\documentclass{article}
\usepackage{FPSpro}
\begin{document}
\title{ 
HIGH PRECISION COSMOLOGY
  }
\author{
  Amedeo Balbi      \\
  {\em Dipartimento di Fisica - Universit\`a di Roma `Tor Vergata' } \\
  {\em INFN - Sezione di Roma II  }\\
  {\em Via della Ricerca Scientifica 1, I-00133 Roma, Italy}
  }
\maketitle
\baselineskip=11.6pt
\begin{abstract}
  I review the current status of cosmology as emerging from recent
  observations of cosmic microwave background anisotropies as well as
  from other sources of cosmological information.
\end{abstract}
\baselineskip=14pt
\section{Introduction}
The widely accepted paradigm for cosmology is the hot Big Bang model.
In this framework, the geometry and evolution of the Universe is
defined by its matter and energy content through general relativity
theory. The Universe is expanding, so that it was hotter and denser at
earlier times. The rate of expansion is quantified by the Hubble
parameter $H$, whose present value $H_0$ is parameterized by the
quantity $h$ as $H_0=100\;h$ km~s$^{-1}$~Mpc$^{-1}$. The amount of
matter and energy in the Universe from different components (baryons,
dark matter, radiation, vacuum energy, etc.) is parameterized by the
quantities $\Omega_{(i)}\equiv\rho_{(i)}/\rho_c$. The critical
density, $\rho_c=1.88\times 10^{-29}\;h^2$~g~cm$^{-3}$, is defined in
such a way that $\Omega\equiv\sum_i \Omega_{(i)}=1$ for a Universe
with flat geometry (while $\Omega<1$ and $\Omega>1$ for open and closed
geometry respectively).

An additional ingredient of the standard cosmological model is {\em
  inflation}\cite{inflation}, a phase of early superluminal expansion
of the Universe required to solve some problems of the Big Bang model.
Inflation makes some well-defined predictions. First of all, the
geometry of the Universe has to be very close to flat. Second, the
structure we observe today in the Universe was produced by
gravitational amplification of primordial density perturbations
generated during inflation, characterized by having a nearly
scale-invariant spectrum and by being Gaussian distributed.

While until recent times the knowledge of the parameters of the
cosmological model was plagued by large uncertainties, the situation
has now dramatically changed.  Cosmology is not a data-starved science
anymore. In the past few years, high-quality observations have fueled
an impressive progress in our understanding of the Universe. We have
entered the epoch of high precision cosmology.

Recent results from observation of the CMB temperature anisotropy have
allowed us to constrain most cosmological parameters to unprecedented
accuracy, giving for the first time a robust determination of the
total energy density (and in turn of the geometry) of the Universe. In
addition, a whole set of new observations of the large-scale structure
properties of the Universe have put the determination of the mean
matter density in the Universe on a firm ground. Finally, measurements
of distant Type Ia Supernovae have recently provided evidence that the
Universe has just entered a phase of accelerated expansion.  In the
following I will review the emerging scenario, giving particular
emphasis to CMB as a cosmological probe.
\section{Cosmology with the Cosmic Microwave Background}
The Cosmic Microwave Background (CMB) is a snapshot of the infant
Universe, when it was just about $300\:000$ years old. According to the
standard Big Bang model, before that epoch the temperature in the
Universe was so high that no neutral atom could stably exist. The
Universe was basically a plasma of mainly free electrons and protons,
kept in equilibrium with photons by frequent Thomson scattering.
Later, the Universe cooled down as a result of the expansion, and
neutral atoms began to form. The photons could then decouple from the
matter and travel freely, being finally observed today as an almost
uniform background.  The fact that the CMB was indeed found to have a
black-body spectrum (a clear signature of the early period of
matter-radiation equilibrium) with an astonishing precision\cite{firas}
is one of the big successes of the Big Bang model.

Since the distribution of the CMB photons reflects that of matter at
the time of decoupling, any inhomogeneities in the matter density
(needed to seed structure formation by gravitational instability) must
leave an imprint as fluctuations of the CMB temperature. The presence
of these {\em CMB temperature anisotropies} was first detected by
NASA's COBE satellite in the early 90's\cite{dmr}. The fact that the
level of anisotropy is very small (about a part in one thousand,
corresponding to temperature fluctuations of some tens of $\mu K$)
simplifies the task of making theoretical prediction of the anisotropy
pattern, since linear perturbation theory can be applied.

The bulk of the cosmological information encoded in the anisotropy
pattern is concentrated at angular scales smaller than about 1 degree
on the sky, corresponding to perturbations that were inside the
horizon (i.e.\ in causal contact) before decoupling. On these scales,
physical processes in the early Universe were able to leave their
imprint on the CMB. For this reason, over the last decade a large
number of ground-based and balloon-borne experiments performed
observations of the fine-structure pattern of the anisotropy.

The observed temperature fluctuation in a given direction of the sky
can be expanded in spherical harmonics:
\begin{equation}
{\Delta T\over T}(\theta,\phi)=
\sum_{l m} a_{l m} Y_{l m}(\theta,\phi).
\end{equation}
The coefficients $C_l \equiv\langle\vert a_{l m}\vert^2\rangle$ define
the {\em angular power spectrum} of the CMB anisotropy\footnote{The
  symbol $\langle\cdot\rangle$ represents an average over the
  statistical ensemble. Since we can only observe one realization of
  the ensemble --- our own sky --- we can at best build an un-biased
  estimate of $C_l$ from the observations. This is: $C_l\equiv {1\over
    (2l+1)}\sum_{m=-l}^{l} \vert a_{l m}\vert^2$.}.  Because the
Universe is isotropic on average, the $C_l$'s do not depend on the
azimuthal index $m$.  If the primordial density fluctuations are
Gaussian distributed, the angular power spectrum $C_l$ fully
characterizes the statistics of the temperature anisotropy pattern.
The power spectrum is then the main CMB observable.  Since each $l$ is
related to an angular scale $\theta$ on the sky given approximately by
$l \sim \pi/\theta$, the power spectrum at high $l$'s probes
sub-horizon angular scales at the time of decoupling and carries the
imprint of physical processes which occurred in the early Universe.
Conversely, low $l$'s probe the primordial shape of the power
spectrum\footnote{Of course, neglecting secondary processes which may
  alter the CMB photon distribution after decoupling.}.

The way the shape of the CMB angular power spectrum depends on
cosmology can be understood by simple physical considerations.  Let us
consider a density fluctuation of given physical scale in the
baryon-photon fluid. Let us suppose that the physical scale of the
fluctuation is smaller than the horizon size at decoupling, so that
the inner region of the fluctuation is in causal contact. The
amplitude of perturbation in the baryon component tends to be
amplified by gravitational collapse. However, the radiation pressure
provided by the photons prevents the collapse from happening.  These
competing mechanisms sets up harmonic oscillations in the amplitude of
the perturbation. Since the amount of resistance to compression is
quantified by the sound velocity in the fluid, this oscillations are
called {\em acoustic}. When the photons decouple from matter,
perturbations having different physical scale are caught in a different
stage of oscillation and then have a different amplitude.  The CMB
photons we receive today carry this phase information as fluctuations
in their temperature at different angular scales. This reflects in a
series of harmonic acoustic peaks in the CMB angular power spectrum.

\begin{figure}[t]
  \vspace{5.5cm}
  \includegraphics{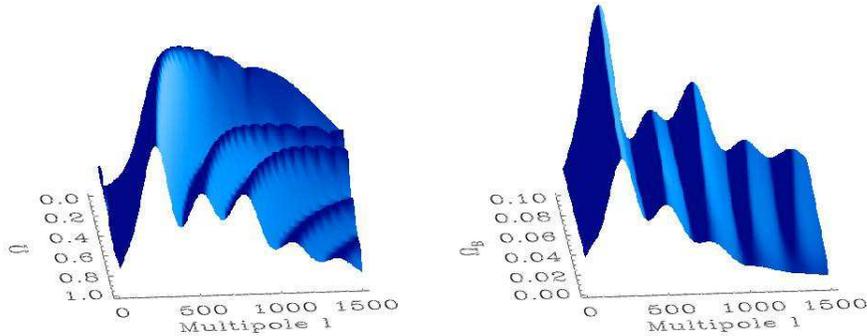}
  \caption{\it The effect of cosmological parameters on the peak 
    structure of the CMB angular power spectrum. On the left, the
    effect of varying the total energy density while keeping all the
    other parameters fixed. On the right, the effect of varying the
    baryon density.
     \label{ClVSpar}}
\end{figure}

For a given initial distribution of density perturbations in the early
Universe, the height of the acoustic peaks is mostly affected by the
amount of matter in the Universe. If we enhance the baryon content of
the Universe, keeping fixed all the other components, the compression
stage of the fluid is more effective, increasing the amplitude of
fluctuations at decoupling.  Then, the relative height of the peaks in
the CMB power spectrum represents a good indicator of the density of
baryonic matter in the Universe.  On the other hand, the position of
the peaks depends on the way a certain physical scale at decoupling is
mapped into an angular dimension on the sky. This is quantified, in a
given cosmological model, by the so called {\em angular diameter
  distance relation}.  This relation mainly depends on the geometry of
the Universe: in an open Universe, a certain physical scale at
decoupling is seen today under a smaller angle than in a flat
Universe. So, the position of the peaks in the CMB angular power
spectrum is a good indicator of the geometrical properties of the
Universe.  The dependence of the CMB angular power spectrum on the
geometry of the Universe and on the baryon density is shown in
Figure~\ref{ClVSpar}.
\section{Constraints on Cosmological Parameters from the CMB}
\begin{figure}[t]
  \vspace{9cm}
  \includegraphics{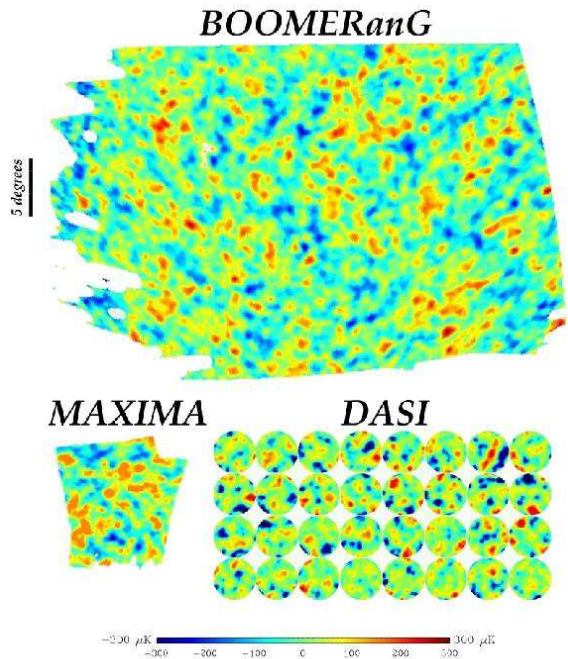}
  \caption{\it Maps of the CMB temperature anisotropy produced 
    by the BOOMERanG, MAXIMA and DASI experiments. 
     \label{maps}}
\end{figure}

The quality of CMB observations has considerably improved in recent
times.  The balloon-borne observations carried on by the
BOOMERanG\cite{boom} and MAXIMA\cite{max} teams (from Antarctica and
from Texas, respectively) have produced the first images of the
fine-scale pattern of CMB temperature anisotropy.  The CMB map from
BOOMERanG covers a 1800 square degrees patch of the southern sky.
MAXIMA mapped a 124 square degrees patch of the northern sky. More
recently, the DASI\cite{dasi} team released new maps over 32 sky
fields of 3.4 degrees in diameter, obtained using ground-based
interferometry from Antarctica.  The kind of spatial features observed
by these three independent experiments in different sky regions looks
quite similar (see Figure~\ref{maps}).

\begin{figure}[t]
  \vspace{8cm} \includegraphics{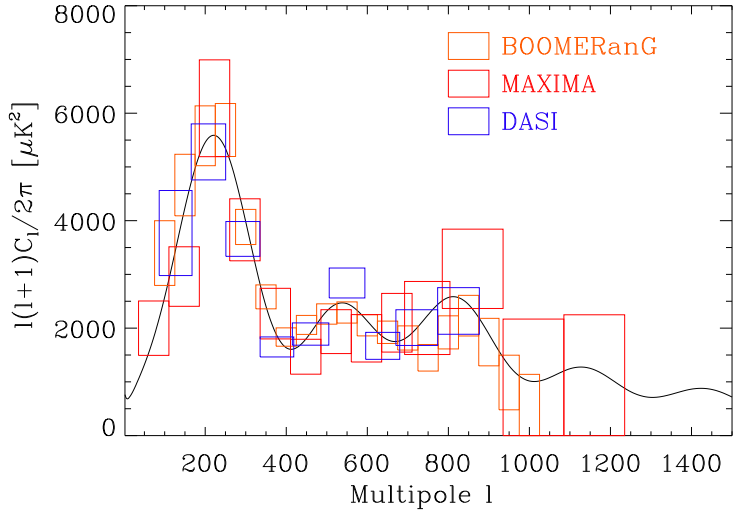}
  \caption{\it Measurements of the CMB angular power spectrum from 
    BOOMERanG, MAXIMA and DASI. The continuous line a reference
    theoretical model for a flat cosmology.
     \label{cl}}
\end{figure}

From these observations, estimates of the CMB angular power spectrum
have been obtained over a large range of multipoles ($20\leq l \leq
1200$; see Figure~\ref{cl}). The power spectra measured by BOOMERanG,
MAXIMA and DASI are in remarkable agreement and show unambiguously the
presence of a sharp peak in the region $180 \leq l \leq 220$, as
well as evidence of excess power at higher $l$'s, consistent with the
presence of a second and third peak.

Likelihood analyses of these power spectrum measurements have been
performed by each team to set constraints on the value of cosmological
parameters. They agree about the fact that the CMB data strongly favor
a Universe with flat geometry, and with scale-invariant primordial
density fluctuations: the inflationary scenario brilliantly passed two
important tests. Furthermore, the baryon density derived from the CMB
is in striking agreement with the value resulting from comparing the
measured primordial light elements abundances with the big bang
nucleosynthesis (BBN) predictions: $\Omega_b h^2=0.020\pm
0.002$\cite{bbn}. This is an important indication of the
self-consistency of our cosmological model, since the CMB and BBN
values for the baryon density are obtained using entirely different
methodology and observations.

\section{The Concordance Model}
The success of the CMB in giving us a reliable estimate of the total
energy density of the Universe leaves us with the problem of finding
out which is the contribution from different components to the
critical density.  Measuring the mean mass density of the Universe
with traditional cosmological observations has always been a difficult
task. Large enough samples have to be observed in order to be
representative of the whole Universe. Furthermore, the distribution of
matter cannot be directly deduced from that of light. However, the
matter density is currently constrained by a number of independent and
consistent observations (baryon-to-total mass ratio in clusters of
galaxies, peculiar velocities and bulk flows, redshift surveys) to be
roughly 1/3 of the total energy density ($\Omega_M=0.33\pm
0.04$\cite{matter}). Where does the rest of critical density comes
from?

Observations of distant type Ia supernovae\cite{sn1a} recently allowed
to probe the classic Hubble diagram up to very high redshifts.  The
surprising result was that, contrarily to expectations, the Universe
is speeding up rather than slowing down. The fact that we are now
entering a phase of cosmic acceleration has been explained with the
presence of a smooth, negative-pressure component, which has been
named {\em dark energy}. The best candidate for dark energy is a
cosmological constant, or vacuum energy, i.e. the vacuum expectation
value of some fundamental scalar field.

\begin{figure}[t]
  \vspace{6cm}
  \includegraphics{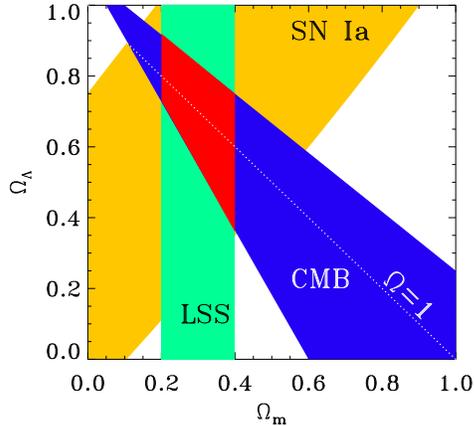}
  \caption{\it Likelihood contours (95\% confidence level) from CMB, supernovae and large-scale structure observations. 
     \label{concordance}}
\end{figure}

Cosmological models with flat geometry but different amount of vacuum
energy have almost the same angular diameter distance relation. This
makes the CMB angular power spectrum basically unable to distinguish
which fraction of the critical density is provided by matter and which
by the vacuum energy. However, when we look at the constraints in the
$\Omega_M$---$\Omega_\Lambda$ plane coming from the CMB, the
observation of large-scale structure (LSS) and type Ia supernovae (SN
Ia) an interesting picture emerges (see Figure~\ref{concordance}). The
CMB and the LSS suggest that 2/3 of the critical density must be
provided by vacuum energy. The CMB and the SN Ia get to the same
conclusion. The three constraints taken together identify a {\em
  concordance} region in the parameter space where $\Omega_M\sim 1/3$,
$\Omega_\Lambda\sim 2/3$, and $\Omega=\Omega_M+\Omega_\Lambda\sim 1$.
The fact that three independent and different kinds of observation,
each probing a different epoch of the cosmic evolution and different
physical processes, have converged to give a coherent picture is a big
success of cosmology.
\section{Future Prospects}
While a consistent and reliable picture of the Universe is emerging,
there are still open questions. One of the most puzzling aspects is
the nature of the dark energy which seems to be the main contribution
to the density of the Universe.  The vacuum energy estimated from
quantum field theory (as vacuum expectation value of some fundamental
quantum field) is $10^{122}$ to $10^ {55}$ times larger than the
observed one, which leads to an extreme {\em fine-tuning} problem.
Furthermore, vacuum energy is dominating the cosmic expansion right
now, which seems to make the present epoch a very special one in the
evolution of the Universe ({\em coincidence} problem). This problems are
mitigated in the so-called {\em quintessence} models, where the scalar
field responsible for the vacuum energy contribution is evolving
through an equation that admits {\em tracking solutions}: large set of
initial conditions result in the same vacuum energy at present.
Attempts to use current CMB data to investigate the nature of dark
energy have recently been made\cite{quintessence}.

Future CMB missions from space will shed more light on this and other
open problems. The NASA's MAP
mission\footnote{http://map.gsfc.nasa.gov/} is currently operating and
will soon produce full sky maps of the CMB sky at high angular
resolution. In 2007 the ESA's Planck
satellite\footnote{http://astro.estec.esa.nl/Planck} will measure CMB
temperature and polarization over the full sky with unprecedented
angular resolution and instrumental sensitivity, reaching the
theoretical limit in the power spectrum measurement over a large range
of multipoles ($2\leq l \leq 3000$). This observations, together with
other sources of information (most notably further supernovae
measurements from space such as those expected from the SNAP
satellite\footnote{http://snap.lbl.gov} and redshift surveys such as
SDSS\footnote{http://www.sdss.org}) will further strengthen our
understanding of the Universe.

\section{Acknowledgements}
It is a pleasure to thank the organizers for the invitation and for
the enjoyable and stimulating environment.

\begin{thebibliography}{99}
  
\bibitem{inflation} Guth, A.H., {\it Phys Rev} {\bf D23}, 347 (1981);
  for a recent pedagogical exposition see Albrecht, A.,
  astro-ph/0007247 (2000).

\bibitem{firas} Fixsen, D.J., et al., {\it ApJ}, {\bf 473}, 576 (1996).
  
\bibitem{dmr} Smoot, G.F., et al., {\it ApJ}, {\bf 396}, L1 (1992).

\bibitem{boom} de Bernardis, P., et al., {\it Nature}, {\bf 404}, 955
  (2000); Lange, A.E., et al., {\it Phys. Rev.} {\bf D63} 042001
  (2001); de Bernardis, P., et al., {\it ApJ}, in press,
  astro-ph/0105296 (2001); Netterfield, C.B., et al., {\it ApJ}, in
  press, astro-ph/0104460 (2001).

\bibitem{max} Balbi, A., et al., {\it ApJL}, {\bf 545}, L1 (2000);
  Hanany, S., et al., {\it ApJL}, {\bf 545}, L5 (2000); Lee, A.T., et
  al., {\it ApJL}, in press, astro-ph/0104459 (2001); Stompor, R., et
  al., {\it ApJL}, in press, astro-ph/0105062 (2001).
  
\bibitem{dasi} Halverson, N.W., et al., {\it ApJ}, in press,
  astro-ph/0104489 (2001); Pryke, C., et al., {\it ApJ}, in press,
  astro-ph/0104490 (2001).

\bibitem{bbn} Burles, S., Nollett, K.M., Truran, J. N., \& Turner, M. S.
{\it Phys. Rev. Lett.}, {\bf 82}, 4176 (1999).

\bibitem{matter} Turner, M.S. submitted to {\it ApJ}, astro-ph/0106035
  (2001).
  
\bibitem{sn1a} Perlmutter, S., et al., {\it ApJ}, {\bf 517}, 565
  (1999); Riess, A.G., et al., {\it AJ}, {\bf 116}, 1009 (1998).
  
\bibitem{quintessence} Balbi, A., Baccigalupi, C., Matarrese, S., Perrotta,
  F.  \& Vittorio, N., {\it ApJ}, {\bf 547}, L89 (2000); Baccigalupi,
  C., Balbi, A., Perrotta, F., Matarrese, S. \& Vittorio, N.,
  submitted to {\it Phys. Rev.  D}, astro-ph/0109097 (2001)


\end{thebibliography}
\end{document}